\documentclass[12pt]{article}
\usepackage{amsmath,amssymb,amsfonts,amsthm}
\usepackage[english]{babel}
\usepackage[cp1251]{inputenc}
\usepackage[toc]{appendix}

\newcommand{\be}{\begin{equation}}
\newcommand{\ee}{\end{equation}}
\newcommand{\bea}{\begin{eqnarray}}
\newcommand{\eea}{\end{eqnarray}}

\newcommand{\lb}{\label}
\newcommand{\nn}{\nonumber \\}

\usepackage[left=1in,right=1in,top=1in,bottom=1in,bindingoffset=0cm]{geometry}
\numberwithin{equation}{section}

\begin{document}
\begin{center}
{\Large\bf ${\cal N}=2$ Higher Spins by Harmonic Superspace Methods}
\vspace{0.8cm}

{\bf Evgeny Ivanov}  \vspace{0.4cm}

Bogoliubov Laboratory of Theoretical Physics,
JINR, \\
141980, Dubna, Moscow Region, Russia\\
{\tt eivanov@theor.jinr.ru}

\end{center}
\vspace{0.5cm}
\begin{flushright}
{\it To the memory of George  Poghosyan{\qquad\quad}}
\end{flushright}
\vspace{0.5cm}

\begin{abstract}
\noindent Harmonic ${\cal N}=2$ superspace was discovered  in 1984 as the powerful tool of the geometric superfield off-shell description of
${\cal N}=2, 4D$ supersymmetric field theories with the maximal spins 1, 2, and 1/2 (${\cal N}=2$ Yang-Mills theories, supergravity and matter hypermultiplets).
My talk is a brief account of the basic achievements of the harmonic methods, including the newest
applications in  ${\cal N}=2$ theories of higher spins.
\end{abstract}

\setcounter{equation}{0}
\section{Supersymmetry and superfields}
Supersymmetry, despite lacking experimental confirmations, is in the heart of the modern of mathematical and quantum physics. It allowed to construct a lot of new theories
  with remarkable and surprising features: supergravities, superstrings, superbranes, ${\cal N}=4$ super
  Yang-Mills theory (the first example of
  the ultraviolet-finite quantum field theory), etc. It also exhibited unexpected relations between these theories, e.g.,
  the ``gravity/gauge'' duality.

The natural approach to supersymmetric theories is the superfield methods. The generalization of Minkowski space $x^m$ to supersymmetry is ${\cal N}$ {\it extended Minkowski superspace}
\bea
{\cal M}^{(4|4{\cal N})} = \left( x^m\,,\;\theta^\alpha_i\,, \;\bar\theta^{\dot\alpha\,i}\right), \; i =1, \ldots, {\cal N}\,,\nonumber
\eea
where $\theta^\alpha_i\,,\bar\theta^{\dot\alpha\,i}$  are anticommuting Grassmann coordinates, $\{\theta, \theta\}, \{\theta, \bar\theta\}= 0$.
The supersymmetric theories are adequately formulated off shell in terms of superfields defined on various superspaces (see, e.g., \cite{1001,BK} and refs. therein).\\

\noindent{\it ${\cal N}=1, 4D$ chirality as the simplest Grassmann analyticity.} ${\cal N}=1, 4D$ supersymmetry is an extension of the Poincar\'e symmetry by spinor generators $Q_\alpha, \bar Q_{\dot\alpha}$
\bea
\{Q_\alpha, \bar Q_{\dot\beta}\} = 4\,P_{\alpha\dot\beta}\,, \, \{Q_\alpha, Q_{\beta}\} =  \{\bar Q_{\dot\alpha}, \bar Q_{\dot\beta}\} = 0\,, \,
[ P_m, Q_\alpha] = [P_m, \bar Q_{\dot\alpha}] = 0\,, \,P_{\alpha\dot\beta} = \frac12 (\sigma^m)_{\alpha\dot\beta} P_m . \nonumber
\eea
${\cal N}=1, 4D$ superspace is an extension of Minkowski space by a doublet of Grassmann anticommuting spinorial coordinates
$\theta^\beta,\bar\theta^{\dot\beta}$,
\bea
x^{\alpha\dot\alpha} \,\Rightarrow \, ( x^{\alpha\dot\alpha} , \theta^\beta, \bar\theta^{\dot\beta}) \,, \, {\theta^\beta}' = \theta^\beta + \epsilon^\beta\,, \,{x^{\alpha\dot\alpha}}'
= {x^{\alpha\dot\alpha}} -2i (\theta^\alpha\bar\epsilon^{\dot\alpha} -\epsilon^\alpha\bar\theta^{\dot\alpha}).\nonumber
\eea
${\cal N}=1, 4D$ superfields  $\Phi(x,\theta, \bar\theta)$ are functions given on ${\cal N}=1$ superspace, with the transformation law,
\bea
\Phi'(x',\theta', \bar\theta') = \Phi(x,\theta, \bar\theta). \nonumber
\eea

There is one very essential distinction between Minkowski space and ${\cal N}=1$ superspace. While the former does not include any coordinate subset
where the whole $4D$ Poincar\'e symmetry could be linearly realized, the latter contains such smaller supermanifolds, ${\cal N}=1, 4D$ chiral
superspaces $(x_L, \theta), (x_R, \bar\theta)$,  with twice as less Grassmann coordinates:
\bea
x_L^{\alpha\dot\beta} = x^{\alpha\dot\beta} + 2i\theta^\alpha\bar\theta^{\dot\beta}\,, \quad \delta  x_L^{\alpha\dot\beta} = -4i\theta^\alpha \bar\epsilon^{\dot\beta}\,,
\quad x_R^{\alpha\dot\beta} = (x_L^{\alpha\dot\beta})^\dagger\,. \nonumber
\eea
The chiral superfields are carriers of the basic matter ${\cal N}=1$ multiplet:
\bea
\varphi(x_L, \theta) = \phi(x_L) + \theta^\alpha\psi_\alpha (x_L) + (\theta)^2 F(x_L)\,,
 S_{free} \sim \int d^4 x  d^2 \theta d^2 \bar\theta \,\varphi(x_L, \theta)\bar\varphi(x_R, \bar\theta)\,. \nonumber
\eea
The chiral superfields can be looked upon as complex general ${\cal N}=1$ superfields subject to the covariant {\it Grassmann analyticity} condition \cite{GrassAnal}.
\bea
\varphi(x_L, \theta) = \Phi_L(x_L, \theta, \bar\theta)\,, \quad \frac{\partial}{\partial \bar\theta^{\dot\gamma}}\Phi_L  = 0\,.  \nonumber
\eea
The same constraint can be rewritten in the basis $(x, \theta, \bar\theta)$ in terms of spinor covariant derivatives
\bea
&&\bar D_{\dot\gamma}\Phi_ L(x, \theta, \bar\theta) = 0\,, \quad  D_\alpha = \frac{\partial}{\partial \theta^\alpha}  + 2i\bar\theta^{\dot\alpha}\partial_{\alpha\dot\alpha}\,,\;
\bar D_{\dot\gamma} = -\frac{\partial}{\partial \bar\theta^{\dot\gamma}}  - 2i`\theta^{\alpha}\partial_{\alpha\dot\gamma}\,, \nonumber \\
&& \{D_{\gamma}, D_{\beta}\} =  \{\bar D_{\dot\gamma}, \bar D_{\dot\beta}\} = 0\,, \quad   \{D_{\gamma}, \bar D_{\dot\beta}\} = -4i \partial_{\gamma\dot\beta}\,. \nonumber
\eea

The vanishing of anticommutators of the same chirality spinor derivatives is just the integrability condition for ${\cal N}=1$ chirality. This chirality underlies
all the gauge and supergravity ${\cal N}=1$ theories: the interacting case just corresponds to replacing all covariant derivatives by the gauge-covariant
ones through adding proper superfield gauge connections
\bea
D_{\gamma} \; \Rightarrow \; {\cal D}_{\gamma} = D_{\gamma} +i {\cal A}_\alpha\,, \,\, \bar D_{\dot\gamma}\; \Rightarrow \; \bar{{\cal D}}_{\dot\gamma} = \bar D_{\dot\gamma}+ i \bar{\cal A}_{\dot\gamma}\,, \,\,
\partial_{\gamma\dot\beta} \; \Rightarrow \;{\cal D}_{\gamma\dot\beta} = \partial_{\gamma\dot\beta} + i {\cal A}_{\gamma\dot\beta}\,,\nonumber
\eea
still preserving the flat integrability constraints
\bea
\{{\cal D}_{\gamma}, {\cal D}_{\beta}\} =  \{\bar{{\cal D}}_{\dot\gamma}, \bar{{\cal D}}_{\dot\beta}\} = 0\,. \nonumber
\eea
The general ${\cal N}=1$ matter is also described by chiral superfields,  implying a general K\"ahler target geometry for bosonic fields \cite{Zumino1979}.

For extended supersymmetries (with few sorts of $Q$ generators) new kinds of Grassmann analyticities (different from chirality) can be defined.
One of them, the harmonic $SU(2)$ analyticity, just forms the basis of the {\it Harmonic Superspace} approach.

\setcounter{equation}{0}
\section{Harmonic superspace}

In four dimensions, the only self-consistent off-shell superfield formalism for ${\cal N}=2$ (and ${\cal N}=3$) theories
is the harmonic superspace approach invented by A. Galperin, E. Ivanov, S. Kalitzin, V. Ogievetsky and E. Sokatchev \cite{1984,1985,Kniga}.

Harmonic ${\cal N}=2$ superspace is defined as the coordinate set:
\bea
Z = (x^m\,,\;\theta^\alpha_i\,, \;\bar\theta^{\dot\alpha\,j}, u^{\pm i}), \quad u^{\pm i} \in SU(2)/U(1),  \; u^{+ i}u^-_i = 1\,. \lb{DefHSS}
\eea
Its most important property is the presence of analytic harmonic ${\cal N}=2$ super-subspace:
\bea
\zeta_A = (x^m_A, \theta^{+ \alpha}, \bar\theta^{+ \dot\alpha}, u^{\pm i}), \; \theta^{+ \alpha, \dot\alpha} := \theta^{\alpha, \dot\alpha i} u^+_i,
\; x^m_A := x^m - 2i\theta^{(i}\sigma^m \bar\theta^{j)}u^+_iu^+_j\,.
   \nonumber
\eea
It is closed under ${\cal N}=2$ supersymmetry. All basic ${\cal N}=2$ superfields are harmonic analytic:
\bea
\underline{\rm SYM}:&& V^{++}(\zeta_A)\,, \;\;  \underline{\rm matter \;hypermultiplets}: \; q^{+}(\zeta_A)\,, \,\bar{q}^{+}(\zeta_A) \nonumber \\
\underline{\rm supergravity}:&& H^{++ m}(\zeta_A)\,,\,H^{++ \hat\alpha +}(\zeta_A)\,, \,H^{++ 5}(\zeta_A)\,, \hat\alpha = (\alpha, \dot\alpha)\,.  \nonumber
\eea
Here the upper indices ``+'' refer to the charge with respect to the denominator of the harmonic coset in \eqref{DefHSS}.\\

\noindent{\it Example: ${\cal N}=2$ spin 1 multiplet}. An instructive example is Abelian ${\cal N}=2$ gauge theory,
\bea
V^{++}(\zeta_A)\,, \quad \delta V^{++} = D^{++}\Lambda (\zeta_A)\,, \; D^{++} = \partial^{++} - 4i\theta^{+\alpha}\bar\theta^{+\dot\alpha}\partial_{\alpha\dot\alpha}\,. \nonumber
\eea
It is possible to impose Wess-Zumino gauge (8 + 8 off-shell degrees of freedom):
\bea
&& V^{++}(\zeta_A) = (\theta^+)^2 \phi + (\bar\theta^+)^2 \bar\phi + 2i\theta^{+\alpha}\bar\theta^{+\dot\alpha} A_{\alpha\dot\alpha} \nonumber \\
&& +\, (\bar\theta^+)^2 \theta^{+\alpha}\psi_\alpha^i u^-_i +
(\theta^+)^2 \bar\theta^{+}_{\dot\alpha}\bar\psi^{\dot\alpha i} u^-_i +(\theta^+)^2(\bar\theta^+)^2 D^{(ik)}u^-_iu^-_k\,. \nonumber
\eea
These component fields constitute an Abelian gauge ${\cal N}=2$ off-shell multiplet.

The invariant superfield action reads:
\bea
&& S \sim \int d^{12}Z\,\, \big(V^{++} V^{--}\big)\,, \; D^{++} V^{--} - D^{--} V^{++} = 0\,, \; \delta V^{--} = D^{--}\Lambda\,, \lb{HCC} \\
&& [D^{++}, D^{--}] = D^0, \quad D^0 V^{\pm\pm} = \pm 2\,V^{\pm\pm}\,. \nonumber
\eea
The second harmonic gauge connection $V^{--}$ is not analytic but it is expressed through $V^{++}$ by the ``harmonic zero-curvature condition'' in \eqref{HCC}.

\section{Supersymmetry and higher spins}
Supersymmetric higher-spin theories provide
 a bridge between superstring theory and low-energy (super)gauge theories.

Free massless bosonic and fermionic higher spin field theories were defined in \cite{Fronsdal1978,  FangFronsdal1978}.

The component approach to  ${\cal N}=1$, $4D$ supersymmetric
free massless higher spin models was developed in \cite{Courtright1979, Vasiliev1980}. The natural tools to deal with supersymmetric theories are off-shell superfield methods.
In the superfield approach the supersymmetry is closed on the off-shell supermultiplets and so is automatically manifest.
The complete off-shell ${\cal N}=1$ superfield Lagrangian formulation of ${\cal N}=1, 4D$
free higher spins was given by S.M.~Kuzenko, {\it et al} \cite{Kuzenko1993, Kuzenko1994}.

An off-shell superfield Lagrangian formulation for higher-spin {\bf extended}
 supersymmetric theories, with all supersymmetries manifest,  was unknown for long
even for free theories. This gap was filled in by I.~Buchbinder, E.~Ivanov, N.~Zaigraev \cite{BIZ2021}.
An off-shell manifestly ${\cal N}=2$ supersymmetric unconstrained formulation of ${\cal N}=2, 4D$ super Fronsdal theory for integer spins was constructed  in the harmonic superspace approach.
Manifestly ${\cal N}=2$ supersymmetric off-shell cubic couplings of ${\cal N}=2, 4D$ higher-spin gauge superfields to the matter hypermultiplets were
further constructed in \cite{BIZ2022, BIZ2023}.

Quite recently, we generalized HSS non-conformal construction to the case of ${\cal N}=2$ superconformal multiplets
and their hypermultiplet couplings \cite{BIZ2024II}.

Our papers just mentioned opened a new domain of applications of the harmonic superspace formalism, that time in ${\cal N}=2$ higher-spin theories.

\section{${\cal N}=2$ spin 2 multiplet}

Analogs of the ${\cal N}=2$ Yang-Mills superfield  $V^{++}(\zeta_A)$ are the following set of analytic gauge potentials:
\bea
&& \Big( h^{++m}(\zeta_A)\,, \;h^{++5}(\zeta_A)\,, \; h^{++\hat{\mu}+}(\zeta_A) \Big),  \quad \hat{\mu} = (\mu\,, \dot{\mu})\,, \nonumber \\
&&\delta_\lambda h^{++m } = {D}^{++} \lambda^m + 2i \big( \lambda^{+\alpha} \sigma^m_{\alpha\dot{\alpha}} \bar{\theta}^{+\dot{\alpha}}
        + \theta^{+\alpha} \sigma^m_{\alpha\dot{\alpha}} \bar{\lambda}^{+\dot{\alpha}}\big)\,, \nonumber \\
&&\delta_\lambda h^{++5} = {D}^{++} \lambda^5 - 2i \big(\lambda^{+{\alpha}} \theta^{+}_{\alpha} - \bar\theta^{+}_{\dot{\alpha}}\bar\lambda^{+\dot{\alpha}}\big),
\delta_\lambda h^{++\hat{\mu}+} = {D}^{++} \lambda^{+\hat{\mu}}\,.  \nonumber
\eea

The Wess-Zumino gauge can be chosen as:
\begin{eqnarray}
&&h^{++m}
       =
        -2i \theta^+\sigma^a \bar{\theta}^+ \Phi^m_a
       +  \big[(\bar{\theta}^+)^2 \theta^+ \psi^{m\,i}u^-_i + c.c.\big]+ \ldots  \nonumber \\
&&h^{++5} =
       -2 i \theta^+ \sigma^a \bar{\theta}^+ C_a + \ldots\,, \quad h^{++\mu+} = \ldots \,, \nonumber
\end{eqnarray}
with the residual gauge freedom:
\begin{eqnarray}
\lambda^m \;\Rightarrow\; a^m(x)\,, \; \lambda^5 \, \Rightarrow \; b(x)\,, \;
\lambda^{\mu+} \;\Rightarrow \; \epsilon^{\mu i}(x) u^+_i + \theta^{+\nu}l_{(\nu}^{\;\;\;\mu)}(x)\,. \nonumber
\end{eqnarray}

The physical fields are $\Phi^m_a, \psi^{m\,i}_\mu, C_a$ (spins ${\bf (2, 3/2, 3/2, 1)}$ on shell). In the ``physical'' gauge:
\bea
\Phi^m_a \sim \Phi_{\beta\dot\beta\alpha\dot\alpha}
\Rightarrow \Phi_{(\beta\alpha)(\dot\beta\dot\alpha)} + \varepsilon_{\alpha\beta}\varepsilon_{\dot\alpha\dot\beta} \Phi\,. \nonumber
\eea

\section{${\cal N}=2$ spin 3 and higher spins}

The spin 3 triad  of analytic gauge superfields is introduced as :
\bea
&& \big\{h^{++(\alpha\beta)(\dot\alpha\dot\beta)} (\zeta)\,, \; h^{++ \alpha\dot\alpha}(\zeta), \; h^{++(\alpha\beta)\dot{\alpha}+}(\zeta), \;
h^{++(\dot\alpha\dot\beta){\alpha}+}(\zeta)\big\}\,, \nonumber \\
&& \delta h^{++(\alpha\beta)(\dot{\alpha}\dot{\beta})} = D^{++} \lambda^{(\alpha\beta)(\dot{\alpha}\dot{\beta})}
       + 2i \big[\lambda^{+(\alpha\beta)(\dot{\alpha}}  \bar{\theta}^{+\dot{\beta})}
        + \theta^{+(\alpha}  \bar{\lambda}^{+\beta)(\dot{\alpha}\dot{\beta})}\big],
        \nonumber \\
       && \delta h^{++\alpha\dot{\alpha}} = D^{++} \lambda^{\alpha\dot{\alpha}} - 2i  \big[\lambda^{+(\alpha\beta)\dot{\alpha}} \theta^{+}_{\beta} +
        \bar\lambda^{+(\dot\alpha\dot\beta){\alpha}} \bar\theta^{+}_{\dot\beta}\big], \nonumber \\
        && \delta h^{++(\alpha\beta)\dot{\alpha}+} = D^{++} \lambda^{+(\alpha\beta)\dot{\alpha}}\,, \; \delta h^{++(\dot{\alpha}\dot{\beta})\alpha+}
        = D^{++} \lambda^{+(\dot{\alpha}\dot{\beta})\alpha}\,. \nonumber
        \eea

The bosonic physical fields in  WZ gauge are collected in
\bea
h^{++(\alpha\beta)(\dot{\alpha}\dot{\beta})}
        =
        -2i \theta^{+\rho} \bar{\theta}^{+\dot{\rho}} \Phi^{(\alpha\beta)(\dot{\alpha}\dot{\beta})}_{\rho\dot{\rho}} + \ldots\, \;
h^{++\alpha\dot{\alpha}} =
        -2i \theta^{+\rho} \bar{\theta}^{+\dot{\rho}} C^{\alpha\dot{\alpha}}_{\rho\dot{\rho}} + \ldots \nonumber
\eea

The physical gauge fields are $\Phi^{(\alpha\beta)(\dot{\alpha}\dot{\beta})}_{\rho\dot{\rho}}$ (spin 3 gauge field),
$C^{\alpha\dot{\alpha}}_{\rho\dot{\rho}}$ (spin 2 gauge field) and $\psi^{(\alpha\beta)(\dot{\alpha}\dot{\beta})i}_\gamma$
(spin 5/2 gauge field). The rest of fields are auxiliary. On shell, we are left with the spin content
$({\bf 3, 5/2, 5/2, 2})$.

The general case with the maximal integer spin ${\bf s}$ is spanned by the analytic gauge potentials
\bea
h^{++\alpha(s-1)\dot\alpha(s-1)}(\zeta), h^{++\alpha(s-2)\dot\alpha(s-2)}(\zeta), h^{++\alpha(s-1)\dot\alpha(s-2)+}(\zeta),
h^{++\dot\alpha(s-1)\alpha(s-2)+}(\zeta), \nonumber
\eea
where $\alpha(s) := (\alpha_1 \ldots \alpha_s), \dot\alpha(s) := (\dot\alpha_1 \ldots \dot\alpha_s)$. The relevant gauge transformations
can also be defined and shown to leave, in the WZ-like gauge, the  physical field multiplet with spins $({\bf s, s-1/2, s-1/2, s-1})$.

The on-shell spin contents of ${\cal N}=2$ higher-spin multiplets can be summarized as
\bea
&&\underline{spin\; 1}: \; 1, \,(1/2)^2,\, (0)^2 \nonumber \\
&&\underline{spin\; 2}: \; 2,\, (3/2)^2,\, 1 \nonumber  \\
&&\underline{spin \;3}: \; 3, \,(5/2)^2, \, 2 \nonumber \\
&& ....... \nonumber \\
&&\underline{spin \;s}: \; s, \, (s-1/2)^2, \, s-1\,. \nonumber
\eea
Each spin enters the direct sum of these multiplets twice, in accord with the general Vasiliev theory of $4D$ higher spins \cite{VasilievBas}.
The off-shell contents of the spin ${\bf s}$ multiplet is described by the generic formula $8[{\bf s}^2 + {\bf (s -1)}^2]_B + 8[{\bf s}^2 + {\bf (s -1)}^2]_F$.

\section{Hypermultiplet couplings}

The construction of interactions in the theory of
higher spins is a very important (albeit difficult) task.

There is an extensive literature on the construction of cubic higher spin interactions (see, e.g., \cite{Bengtsson1983, FradkinMetsaev1991, Metsaev1993, MMRu2011}).

Supersymmetric $\mathcal{N}=1$ generalizations of the bosonic cubic vertices with matter were explored in terms of $\mathcal{N}=1$ superfields
in refs. \cite{BuGaKo, GatesKoutrolikosIBuch, MetsaevI}, and in many other papers.

In \cite{BIZ2022,BIZ2023} we have constructed the off-shell
manifestly $\mathcal{N}=2$ supersymmetric cubic couplings
$(\mathbf{\frac{1}{2}, \frac{1}{2}, s})$  of an arbitrary higher
integer  superspin $\mathbf{s}$ gauge $\mathcal{N}=2$ multiplet to the
hypermultiplet matter in $\mathcal{N}=2, 4D$ harmonic
superspace. In our approach $\mathcal{N}=2$ supersymmetry of cubic vertices is always manifest and off-shell, in contrast,
e.g., to the non-manifest light-cone formulations \cite{MetsaevI}.

The starting point is the ${\cal N}=2$ hypermultiplet off-shell free action:

\begin{equation}
    S = \int  d\zeta^{(-4)}  \; \mathcal{L}^{+4}_{free} = -\int d\zeta^{(-4)}  \; \frac{1}{2} q^{+a} \mathcal{D}^{++} q^+_a, a = 1,2\,. \nonumber
\end{equation}

Analytic gauge potentials for any spin ${\bf s}$ with the correct transformation rules are recovered by proper gauge-covariantization
of the harmonic derivative $\mathcal{D}^{++}$. The simplest option is gauging of $U(1)$,

\begin{eqnarray}
&&\delta q^{+a} = -\lambda_0 J q^{+ a}, \quad J q^{+ a} = i (\tau_3)^a_{\;b} q^{+b}\,,\nonumber \\
&& {\cal D}^{++} \;\Rightarrow \;{\cal D}^{++} + \hat{\cal H}^{++}_{(1)}\,, \quad  \hat{\cal H}^{++}_{(1)}= h^{++}J\,, \nonumber \\
&& \delta_\lambda \hat{\cal H}^{++}_{(1)} = [{\cal D}^{++}, \hat{\Lambda}]\,, \quad \hat{\Lambda}= \lambda J\; \Rightarrow \;
\delta_\lambda h^{++} = {\cal D}^{++}\lambda\,. \nonumber
\eea

In ${\cal N}=2$ supergravity, that is for ${\bf s}=2$,

\begin{eqnarray}
&&    S_{(2)} =  -\int d\zeta^{(-4)}  \; \frac{1}{2} q^{+a} \big(\mathcal{D}^{++} + {\cal H}_{(2)}\big)q^+_a, \quad \delta{\cal H}_{(2)} = [\mathcal{D}^{++}, \hat{\Lambda}_{(2)}],\nonumber \\
&& \quad {\cal H}_{(2)}= h^{++ M}(\zeta)\partial_M, \; \hat{\Lambda}_{(2)}
= \lambda^{M}(\zeta)\partial_M, \; M := (\alpha\dot\beta, 5, \hat{\mu}+)\,. \nonumber
\end{eqnarray}

For higher ${\bf s}$ everything goes analogously. For ${\bf s}=3$:
\begin{eqnarray}
&&    S_{(3)} =  -\int d\zeta^{(-4)}  \; \frac{1}{2} q^{+a} \big(\mathcal{D}^{++} + {\cal H}_{(3)}J \big)q^+_a, \nonumber \\
&&\delta{\cal H}_{(3)} = [\mathcal{D}^{++}, \hat{\Lambda}_{(3)}], \quad {\cal H}_{(3)}
= h^{++\alpha\dot\alpha\,M}(\zeta)\partial_M\partial_{\alpha\dot\alpha}, \quad \hat{\Lambda}_{(3)}
= \lambda^{\alpha\dot\alpha\,M}(\zeta)\partial_M\partial_{\alpha\dot\alpha}\,. \nonumber
\end{eqnarray}

\section{Superconformal couplings}

Free conformal higher-spin actions in $4D$ Minkowski space were
pioneered by Fradkin and Tseytlin \cite{FradkinTseytlin1985} and Fradkin and Linetsky \cite{FradkinLinetsky1989, FradkinLinetsky1991}. Since then,
a lot of works on (super)conformal higher spins followed (see, e.g., \cite{Segal2003, ManvelyanMkrtchyanRuehl2010}).

(Super)conformal higher-spin theories are considered as a basis for all other types of higher-spin models. Non-conformal ones follow from the superconformal ones through couplings
to the {\bf superfield compensators}.

In \cite{BIZ2024II} we extended the  off-shell $\mathcal{N}=2, 4D$ higher spins and their hypermultiplet couplings to the superconformal case.

$\mathcal{N}=2, 4D$ superconformal algebra (SCA) preserves harmonic analyticity and is a closure of the rigid $\mathcal{N}=2$ supersymmetry and special conformal symmetry \cite{Kniga}
\bea && \delta_\epsilon \theta^{+ \hat{\alpha}} = \epsilon^{\hat{\alpha}i}u^+_i\,, \; \delta_\epsilon x^{\alpha\dot\alpha} =
-4i \left( \epsilon^{\alpha i} \bar{\theta}^{+\dot{\alpha}} + \theta^{+\alpha} \bar{\epsilon}^{\dot{\alpha}i} \right) u^-_i\,, \hat{\alpha} = (\alpha, \dot\alpha) \,,\nonumber \\
&&\delta_k \theta^{+ {\alpha}} = x^{\alpha\dot\beta}k_{\beta\dot\beta}\theta^{\hat{\beta}}\,, \;\delta_k x^{\alpha\dot\alpha} = x^{\rho\dot\alpha}k_{\rho\dot\rho}x^{\dot\rho\alpha}\,, \; \delta_k u^{+ i} =
(4i\theta^{+\alpha}\bar\theta^{+\dot\alpha}k_{\alpha\dot\alpha}) u^{- i}\,. \nonumber \eea

What about the conformal properties of various analytic higher-spin potentials? No problems with the spin ${\bf 1}$ potential $V^{++}$:
\bea \delta_{sc} V^{++} = - \hat\Lambda_{sc}V^{++}\,, \quad \hat\Lambda_{sc}:= \lambda^{\alpha\dot\alpha}_{sc}\partial_{\alpha\dot\alpha} + \lambda^{\hat\alpha+}_{sc}\partial_{\hat\alpha+} +
\lambda^{++}_{sc}\partial^{--} \,.\nonumber \eea

The cubic vertex $\sim q^{+a}V^{++}Jq^+_a$ is invariant up to total derivative if
\bea
\delta_{sc} q^{+a}  = -\hat\Lambda_{sc} q^{+a} - \frac12 \Omega q^{+a}\,, \quad \Omega :=
    (-1)^{P(M)} \partial_M \lambda^M \,.\nonumber
\eea
Situation gets more complicated for ${\bf s}\geq 2$. Requiring ${\cal N}=2$ gauge potentials for ${\bf s}=2$ to be closed under
${\cal N}=2$ SCA necessarily leads to
\bea
&& {\cal D}^{++} \rightarrow {\cal D}^{++} + \kappa_2 \hat{\mathcal{H}}^{++}_{(s=2)}\,, \nonumber \\
&& \hat{\mathcal{H}}^{++}_{(s=2)} : = h^{++M} \partial_M
    =
        h^{++\alpha\dot{\alpha}}\partial_{\alpha\dot{\alpha}}
        +
    h^{++\alpha+}\partial^-_\alpha
    +
    h^{++\dot{\alpha}+} \partial^-_{\dot{\alpha}}
    +
    h^{(+4)}\partial^{--} \nonumber \\
&&     \delta_{k_{\alpha\dot{\alpha}}} h^{(+4)} = - \hat{\Lambda} h^{(+4)}
    +
    4i h^{++\alpha+} \bar{\theta}^{+\dot{\alpha}} k_{\alpha\dot{\alpha}}
    +
    4i \theta^{+\alpha} h^{++\dot{\alpha}+}
    k_{\alpha\dot{\alpha}}\,.\nonumber
\eea
It is impossible to avoid introducing the extra potential $h^{(+4)}$ for ensuring conformal covariance. This extended set of potentials (with the properly defined gauge freedom)
embodies ${\cal N}=2$ {\bf Weyl multiplet} (${\cal N}=2$ conformal SG gauge multiplet).

For ${\bf s}\geq 3$ the gauge-covariantization of the free $q^{+ a}$ action requires adding the gauge superfield differential operators
of rank ${\bf s -1}$ in $\partial_M$,
\bea {\cal D}^{++} \rightarrow {\cal D}^{++} + \kappa_s \hat{\mathcal{H}}^{++}_{(s)}(J)^{P(s)}\,, \quad P(s) = \frac{1 + (-1)^{s-1}}{2} \,.\nonumber
\eea

For ${\bf s}=3$:
\bea \hat{\cal H}_{(s=3)}= h^{++MN}\partial_N\partial_M + h^{++}, \quad  h^{++MN} = (-1)^{P(M)P(N)} h^{++ NM}\,.\nonumber
\eea
${\cal N}=2$ SCA mixes different entries of $h^{++MN}$, so we need to take into account all these entries, as distinct from the non-conformal case where it was enough to consider, e.g.,
$h^{++\alpha\dot\alpha M}$.

The spin ${\bf 3}$ gauge transformations of $q^{+a}$ and $h^{++MN}$
leaving invariant the action $\sim q^{+a}( D^{++} + \kappa_3\hat{\cal H}_{(s=3)})q^+_a$  are
\bea && \delta_\lambda^{(s=3)} q^{+a}
    = -\frac{\kappa_3}{2}\{\hat{\Lambda}^{M} + \frac12 \Omega^{M}, \partial_{M} \}_{AGB} J q^{+a}
    \equiv -\kappa_3 \hat{{\cal U}}_{(s=3)}Jq^{+a}\,, \nonumber \\
&& \delta^{(s=3)}_\lambda \hat{\mathcal{H}}^{++}_{(s=3)} =
    \left[\mathcal{D}^{++}, \hat{{\cal U}}_{(s=3)}\right], \nonumber \\
&& \hat{\Lambda}^M := \sum\limits_{N\leq M} \lambda^{MN}\partial_N\,, \; \Omega^M := \sum\limits_{N\leq M} (-1)^{\left[P(N) + 1\right] P(M)}\partial_N \lambda^{NM},\nonumber \\
&&\{F_1, F_2\}_{AGB} := [F_1, F_2], \quad \{B_1, B_2\}_{AGB} :=\{B_1, B_2\} \,.\nonumber
\eea
All the  potentials, excepting $h^{++\alpha\dot{\alpha}M}$,
can be put equal to zero using the original extensive  gauge freedom:
\bea
    S^{(s=3)}_{int|fixed}
    =
    -\frac{\kappa_3}{2} \int d\zeta^{(-4)}\; q^{+a} h^{++\alpha\dot{\alpha}M} \partial_M \partial_{\alpha\dot{\alpha}} J q^+_a\,.
\eea

Using the linearized gauge transformations of $h^{++\alpha\dot\alpha M}$,
\begin{eqnarray} &&\delta_\lambda h^{++(\alpha\beta)(\dot{\alpha}\dot{\beta})} =      \mathcal{D}^{++}\lambda^{(\alpha\beta)(\dot{\alpha}\dot{\beta})}
        + 4i \lambda^{+(\alpha\beta)(\dot{\alpha}} \bar{\theta}^{+\dot{\beta})}
        + 4i \theta^{+(\alpha} \bar{\lambda}^{+\beta)(\dot{\alpha}\dot{\beta})},
     \nonumber  \\
  &&      \delta_\lambda h^{++(\alpha\beta)\dot{\alpha}+} =  \mathcal{D}^{++}\lambda^{+(\alpha\beta)\dot{\alpha}} - \lambda^{++(\alpha\dot{\alpha}} \theta^{+\beta)},
       \nonumber\\
     &&   \delta_\lambda h^{++(\dot{\alpha}\dot{\beta})\alpha+} = \mathcal{D}^{++}\lambda^{+(\dot{\alpha}\dot{\beta})\alpha} - \lambda^{++\alpha(\dot{\alpha}} \bar{\theta}^{+\dot{\beta})},
       \nonumber \\
      &&  \delta_\lambda h^{(+4)\alpha\dot{\alpha}} = \mathcal{D}^{++} \lambda^{++\alpha\dot{\alpha}}
        - 4 i \bar{\theta}^{+\dot{\alpha}} \lambda^{+\alpha++}
        +
        4i \theta^{+\alpha} \lambda^{+\dot{\alpha}++}, \nonumber
\end{eqnarray}
we can find WZ gauge for the spin 3 gauge supermultiplet
\begin{eqnarray} && h^{++(\alpha\beta)(\dot{\alpha}\dot{\beta})}
            =
            -4i \theta^{+\rho}\bar{\theta}^{+\dot{\rho}} \Phi_{\rho\dot{\rho}}^{(\alpha\beta)(\dot{\alpha}\dot{\beta})}
            +
            (\bar{\theta}^+)^2 \theta^+ \psi^{(\alpha\beta)(\dot{\alpha}\dot{\beta})i} u_i^-
           \nonumber \\
            &&\;\;\;\;\;\;\;\;\;\;\;\;\;\;\;\;+
            (\theta^+)^2 \bar{\theta}^{+} \bar{\psi}^{(\alpha\beta)(\dot{\alpha}\dot{\beta})i} u_i^-
            +
            (\theta^+)^2 (\bar{\theta}^+)^2 V^{(\alpha\beta)(\dot{\alpha}\dot{\beta})ij}u^-_i u^-_j\,,\nonumber \\
          &&  h^{++(\alpha\beta)\dot{\alpha}+}
            = (\theta^{+})^2 \bar{\theta}^{+}_{\dot{\nu}} P^{(\alpha\beta)(\dot{\alpha}\dot{\nu})}
            +
            (\bar{\theta}^{+})^2 \theta^{+}_{\nu} T^{(\alpha\beta\nu)\dot{\alpha}}_{}
            +
            (\theta^{+})^4 \chi^{(\alpha\beta)\dot{\alpha}i}u_i^-\,,
            \nonumber \\
&& h^{(+4)\alpha\dot{\alpha}} \;\quad =(\theta^+)^2(\bar{\theta}^{+})^2 D^{\alpha\dot\alpha}\,. \nonumber
\end{eqnarray}

 In its {\bf bosonic sector}: the spin ${\bf s}=3$  gauge field, $SU(2)$ triplet of conformal gravitons,
singlet conformal graviton, spin $1$ gauge field and non-standard field which gauges self-dual two-form symmetry:
\bea
    \Phi^{(\alpha\beta\rho)(\dot{\alpha}\dot{\beta}\dot\rho)},\;V^{(\alpha\beta)(\dot{\alpha}\dot{\beta})(ij)},\; P^{(\alpha\beta)(\dot{\alpha}\dot{\nu})}, \;
    D^{\alpha\dot{\alpha}},\;T^{(\alpha\beta\gamma)\dot{\alpha}}\,.\nonumber
\eea
In its {\bf fermionic sector}: conformal spin $5/2$ and spin $3/2$ gauge fields:
\bea
\psi^{(\alpha\beta\rho)(\dot{\alpha}\dot{\beta})i}, \quad \chi^{(\alpha\beta)\dot\alpha i}\,. \nonumber
\eea

They carry total of $ 40 + 40$ off-shell degrees of freedom. Starting from ${\bf s}=3$,
all the component fields are gauge fields, no auxiliary fields are present.

The sum of conformal spin $2$ and spin $3$ actions,
\begin{equation}
    S = - \frac{1}{2} \int d\zeta^{(-4)} \, q^{+a} \left( \mathcal{D}^{++} + \kappa_2 \hat{\mathcal{H}}^{++}_{(s=2)}
    + \kappa_3 \hat{\mathcal{H}}^{++}_{(s=3)}J \right) q^+_a \nonumber
\end{equation}
is invariant with respect to the (properly modified) spin $\mathbf{3}$ transformations to the leading order in  $\kappa_3$ and to any order in $\kappa_2$.
 Thus the cubic
vertex $(\mathbf{3}, \mathbf{\tfrac{1}{2}}, \mathbf{\tfrac{1}{2}})$ is invariant under the gauge transformations of conformal $\mathcal{N}=2$ SG
and we obtain the superconformal vertex of the spin $\mathbf{3}$ supermultiplet on {\it generic}  ${\cal N}=2$
Weyl  SG background.

The whole consideration can be generalized to the general integer higher-spin ${\bf s}$ case. It involves $ 8(2{\bf s} -1)_B + 8(2{\bf s}-1)_F$ off-shell degrees of freedom.

\section{Fully consistent higher-spin hypermultiplet coupling}

The superconformal cubic vertices $(\mathbf{s}, \mathbf{\frac{1}{2}}, \mathbf{\frac{1}{2}})$ are consistent
to the leading order in the higher-spin analogs of Einstein constant. These can be made invariant with respect to gauge transformations
of the whole tower of the higher-spin ${\cal N}=2$ gauge superfields.

To summarize the procedure exposed in \cite{BIZ2024II}, we introduce an analytic differential operator involving {\bf all} integer higher spins:
\begin{equation}
    \hat{\mathcal{H}}^{++} := \sum_{s=1}^{\infty}  \kappa_s \hat{\mathcal{H}}^{++}_{(s)} (J)^{P(s)}\,.\nonumber
\end{equation}

The action of the infinite tower of integer $\mathcal{N}=2$ superconformal higher spins interacting with the hypermultiplet
in an arbitrary $\mathcal{N}=2$ conformal supergravity background then reads:

\begin{equation}
    S_{full} = - \frac{1}{2} \int d\zeta^{(-4)} \,q^{+a}
    \left(\mathcal{D}^{++} + \hat{\mathcal{H}}^{++} \right) q^+_a\,. \nonumber
\end{equation}

Ascribing the proper gauge transformation to $\hat{\mathcal{H}}^{++}$, one can achieve gauge invariance to any order in the couplings constants.
The total hypermultiplet gauge transformation reads
\begin{equation}
    \delta_\lambda q^{+a} = - \hat{\mathcal{U}}_{hyp} q^{+a} = - \sum_{s=1}^{\infty} \kappa_s \, \hat{\mathcal{U}}_{s} \left( J\right)^{P(s)} q^{+a}, \nonumber
\end{equation}
\begin{equation}
    \hat{\mathcal{U}}_{s} q^{+a}: = \sum_{k ={s, s-2, \dots}}   \hat{\mathcal{U}}_{s} ^{(k)} q^{+a}\,. \nonumber
\end{equation}
This transformation acts linearly on the hypermultiplet superfield.

For the set of gauge  superfields we obtain the transformation law:
\begin{equation}
    \delta_\lambda \hat{\mathcal{H}}^{++}
    =
    \left[ \mathcal{D}^{++} +  \hat{\mathcal{H}}^{++},   \hat{\mathcal{U}}_{gauge}\right],
    \qquad
    \hat{\mathcal{U}}_{gauge} : = \sum_{s=1}^\infty \kappa_s \, \hat{\mathcal{U}}_{s}\,. \nonumber
\end{equation}
It mixes different spins, so it is a non-Abelian deformation of the spin $\mathbf{s}$ transformation laws.
In the lowest order, it is reduced to the sum of linearized transformations of {\bf all}
integer spins ${\bf s}\geq 1$.

The invariance under $\mathcal{N}=2$ conformal supergravity transformations is automatic.
So we have constructed the fully consistent gauge-invariant and conformally invariant
interaction of hypermultiplet with an infinite tower of $\mathcal{N}=2$ higher spins
in an arbitrary $\mathcal{N}=2$ conformal supergravity background.

\section{Towards ${\cal N}=2$ AdS background}
It is most interesting to explicitly construct ${\cal N}=2$ higher spins in the AdS background, with the superconformal symmetry $SU(2,2|2)$ being broken
to the AdS supersymmetry $OSp(2|4;R)$. Following a recent work \cite{EN}, we shall deal with embedding of the latter into the former. Another setting \cite{Kuz}
uses the coset superspace formalism as a generalization of the one employed in \cite{ISo} for ${\cal N}=1$ case.

The embedding of ${\cal N}=2$ AdS superalgebra into $SU(2,2|2)$ is specified through the identification \cite{BILS}:
\bea
&&\Psi^i_\alpha = Q^i_\alpha + c^{ik} S_{k\alpha}, \quad \bar{\Psi}^i_{\dot\alpha} = \overline{\Psi^i_\alpha} = \bar{Q}_{\dot\alpha i} + c_{ik}\bar{S}^k_{\dot\alpha}, \nn
&& c^{ik} = c^{ki}\, \qquad \overline{c^{ik}} = c_{ik} = \varepsilon_{il}\varepsilon_{kj} c^{lj}\,. \nonumber
\eea
The $SU(2,2|2)$ superalgebra commutation relations imply for the super AdS generators
\bea
&& \{ \Psi^i_\alpha,  \Psi^k_\beta \} = c^{ik} L_{(\alpha\beta)} +4 i \varepsilon_{\alpha\beta}\varepsilon^{ik} T, \quad T := c_{lm}T^{lm}\,,
\quad [T, \Psi^i_\alpha] \sim c^{ik}\Psi_{k\alpha}\,, \nonumber \\
&& \{ \Psi^i_\alpha,  \bar\Psi_{\dot\beta k}\} = 2\delta^i_k\,R_{\alpha\dot\beta}\,, \;\; R_{\alpha\dot\beta} = P_{\alpha\dot\beta} + \frac12 c^2\, K_{\alpha\dot\beta}\,, \;\;
c^2 := c^{ik}c_{ik} \sim \frac{1}{R^2_{AdS}}, \nonumber \\
&& [R_{\alpha\dot\alpha}, R_{\gamma\dot\gamma}] \sim c^2\big(\varepsilon_{\alpha\gamma} L_{\dot\alpha\dot\gamma} + \varepsilon_{\dot\alpha \dot\gamma} L_{\alpha\gamma} \big),
\quad [R_{\alpha\dot\beta},\Psi^i_\beta]  \sim \varepsilon_{\alpha\beta}\,\bar\Psi^i_{\dot\beta}\; ({\rm and \,c.c.}). \nonumber
\eea

 The first step toward constructing an off-shell ${\cal N}=2$ AdS higher spin theory  is to define the super AdS
invariant Lagrangian of hypermultiplet, such that it respect no full superconformal invariance, but only the super AdS one.

One needs to define the AdS covariant version of the analyticity-preserving harmonic derivative ${\cal D}^{++}$.
The appropriate  ${\cal D}^{++}_{AdS}$ acting on $q^{+a} = (q^+, \tilde{q}^+)$ has the structure
\bea
&&{\cal D}^{++}_{AdS} = \partial^{++} - 4i \hat{\theta}^{+ \alpha} \hat{\bar\theta}^{+ \dot\alpha} \nabla_{\alpha\dot\alpha} + h^{++} \hat{T} + {\cal O}(c)\,,\nn
&& \nabla_{\alpha\dot\alpha} = \big(1 + \frac12c^2 x^2)\partial_{\alpha\dot\alpha}, \quad h^{++} = i|c|\big[ (\hat{\theta}^+)^2 - (\hat{\bar\theta}^+)^2\big] + {\cal O}(c), \nn
&& \hat{T}\, (q^+, \,\tilde{q}^+) = (q^+,\, -\tilde{q}^+)\,, \nonumber
\eea
where $\hat{\theta}^{+}_\alpha, \hat{\bar\theta}^{+}_{\dot \alpha}$ are some redefinitions of the original Grassmann coordinates and ${\cal O}(c)$ stand for terms vanishing
in the limit $c^{ik} \rightarrow 0$.

An extra term $\sim \hat{T}$ in ${\cal D}^{++}_{AdS}$ is necessary for breaking superconformal invariance and it produces
a mass of $q^+$. In the properly defined flat limit this term becomes the central charge extension of flat $D^{++}$ and $\hat{T}$ goes  just into the derivative
$\partial_5$.

More details on the AdS invariant $q^+$ Lagrangians
are given in our recent work with Nikita Zaigraev \cite{EN}.

An interesting new result obtained there  is the analyticity-preserving Weyl transformation of the hypermultiplet Lagrangian.

We start from the free $q^{+}$ action, $S_{free} = -\frac12\int d\zeta^{(-4)}q^{+a}\mathcal{D}^{++} q^+_a$. It
is superconformally invariant and hence invariant under super ${\cal N}=2$ super AdS$_4$ group. Then we make the Weyl-type rescaling of $q^{+}$,
\bea
q^{+a} = G^{\frac12}\,\hat{q}^{+a}\,, \quad G = \frac{\Big(1 + \frac{(c^{+-})^2}{m^2}\Big)}{\Big(1 + \frac{m^2 x^2}{2}\Big)^2} \Big(1 + \theta\,terms\Big)\,, c^{+-} = c^{ik}u^+_iu^-_k\,,\nonumber
\eea
so that $\hat{q}^{+a}$ is a scalar under ${\cal N}=2$ super AdS$_4$ group. The $\hat{q}^{+}$ action takes the form manifestly invariant under this group
\bea
S_{free} = -\frac12\int d\zeta^{(-4)}\,G \,\hat{q}^{+a}\mathcal{D}^{++} \hat{q}^+_a\,, \quad \delta_{osp}\hat{q}^{+a} = 0\,. \nonumber
\eea
The new integration measure  $d \zeta^{(-4)} G$ is invariant  under $OSp(2|4;R)$. So one can add to the Lagrangian  any proper function of $\hat{q}^{+a}$
without breaking of $OSp(2|4;R)$. In particular, one can add an arbitrary ${\cal L}^{+4}(\hat{q}^{+a}, u^-)$ and so get a wide class of the hyper-K\"ahler-like sigma model actions
on the AdS$_4$ background.

\section{Summary and outlook}
The theory of ${\cal N}= 2$ supersymmetric higher spins $s\geq 3$ opens a new promising direction
of applications of the harmonic superspace approach which earlier proved to be indispensable for description of
more conventional ${\cal N}= 2$ theories with maximal spins $s\leq 2$. Once again,
the basic property underlying these new higher-spin theories is the harmonic Grassmann analyticity (all basic gauge potentials
are unconstrained analytic superfields involving an infinite number of degrees of freedom off shell before fixing WZ-type gauges).\\

\noindent{\bf Under way}:

\begin{itemize}
\item The linearized actions of conformal higher-spin ${\cal N}=2$ multiplets (${\cal N}=2$ analogs of the square of Weyl tensor)?

\item Quantization, induced actions,...

\item ${\cal N}=2$ supersymmetric half-integer spins (along the lines of recent ref. \cite{EINZ}).

\item The complete extension to AdS  background? Superconformal compensators? The ${\cal N}=2$ AdS$_4$ supergroup $OSp(2|4;R) \subset SU(2,2|2)$, so the conformal invariance already implies
AdS$_4$ invariance. The latter is extracted as $\eta^{\alpha i} \rightarrow c^{ik} \epsilon^\alpha_k$. It remains to pick up the appropriate  super-compensators (in progress).

\item From the linearized theories to their full nonlinear versions? At present, the latter is known only for $s\leq 2$ (${\cal N}=2$ super Yang - Mills
and ${\cal N}=2$ supergravities). This problem seemingly requires  accounting  for {\bf ALL} higher ${\cal N}=2$ superspins simultaneously. New supergeometries?

\end{itemize}

\section*{Acknowledgements}
I thank the Organizers of the XIII International Symposium on Quantum Theory and Symmetries
for the kind hospitality in Yerevan. I am indebted to Ioseph Buchbinder and Nikita Zaigraev for fruitful collaboration during last years.

\end{document}